\documentclass[twocolumn,floatfix,showpacs,showkeys,preprintnumbers,nofootinbib,superscriptaddress]{revtex4}
\usepackage[utf8]{inputenc}
\usepackage[sort&compress]{natbib}
\usepackage{ulem}
\usepackage{bm}
\usepackage{times}
\usepackage{amssymb,amsbsy,amsmath,amsfonts}
\usepackage{graphicx} 
\usepackage{float}
\usepackage{color}
\usepackage{morefloats}
\usepackage{rotating}
\usepackage{srcltx}
\usepackage{overpic}
\usepackage{slashed}
\usepackage{subfigure}
\usepackage{multirow}
\usepackage{verbatim}
\usepackage{hyperref}
\usepackage{tabularx}

\begin{document}

\title{Systematic studies of $DDKK$ and $D\bar{D}K\bar{K}$ four-hadron molecules}

\author{Ya-Wen Pan}
\affiliation{School of Physics, Beihang University, Beijing 102206, China}

\author{Ming-Zhu Liu}
\affiliation{
Frontiers Science Center for Rare isotopes, Lanzhou University,
Lanzhou 730000, China}
\affiliation{ School of Nuclear Science and Technology, Lanzhou University, Lanzhou 730000, China}

\author{Jun-Xu Lu}\email{ljxwohool@buaa.edu.cn}
\affiliation{School of Physics, Beihang University, Beijing 102206, China}

\author{Li-Sheng Geng}\email{ lisheng.geng@buaa.edu.cn}
\affiliation{School of Physics, Beihang University, Beijing 102206, China}
\affiliation{Beijing Key Laboratory of Advanced Nuclear Materials and Physics, Beihang University, Beijing 102206, China}
\affiliation{Peng Huanwu Collaborative Center for Research and Education, Beihang University, Beijing 100191, China}
\affiliation{Southern Center for Nuclear-Science Theory (SCNT), Institute of Modern Physics, Chinese Academy of Sciences, Huizhou 516000, China}

\begin{abstract}
Assuming that $D_{s0}^{*}(2317)$ is a $DK$ molecular state with a binding energy of 45 MeV, we investigate the existence of four-hadron molecules, $DDKK$ and $D\Bar{D}K\Bar{K}$, with the Gaussian expansion method. Their binding energies are $138\sim155$ MeV and $123\sim163$ MeV below the mass thresholds of $DDKK$ and $D\Bar{D}K\Bar{K}$. The $D\Bar{D}K\Bar{K}$ state has a decay width of  $36\sim54$ MeV due to the complex $K\Bar{K}$ interaction.
Further theoretical studies of and experimental searches for such four-hadron molecules can help deepen the understanding of the nonperturbative strong interaction in a nontrivial way.

\end{abstract}
\maketitle

\section{Introduction}
   
Many studies have been performed on multi-hadron systems. The best-known ones are atomic nuclei and hypernuclei. The deuteron ($np$) is formed by a proton and a neutron with a binding energy of 2.2 MeV~\cite{Greene:1986vb}, the triton ($nnp$) is formed by a proton and two neutrons with a binding energy of 8.5 MeV~\cite{Fujiwara:2002sy}, and the $\alpha$ particle ($nnpp$) is formed by two protons and two neutrons with a binding energy of 28.1 MeV~\cite{Lu:2018pjk}. In addition, as an isotope of helium, the ${}^{3}\rm{He}$ nucleus ($ppn$) is formed by two protons and a neutron. 
Similar bound states composed of different numbers of antikaons and nucleons have also been studied~\cite{Dote:2008zz, Dote:2008xa, Barnea:2012qa, Ikeda:2010tk, Revai:2014twa, Wycech:2008zzb, Shevchenko:2007zz, Kanada-Enyo:2008wsu, Hyodo:2022xhp, Kezerashvili:2021ren}. The $\Lambda(1405)$ can be considered as a quasibound  $\Bar{K}N$ state~\cite{Dalitz:1959dn, Dalitz:1960du}. The $\Bar{K}\Bar{K}N$ system was found to bind with a binding energy of $10\sim33$ MeV~\cite{Kanada-Enyo:2008wsu}. In  Ref.~\cite{Hyodo:2022xhp}, the authors studied multi-hadron states composed of an antikaon and a varying number of nucleons. They found that the binding energies of $\Bar{K}NN$,  $\Bar{K}NNN$, $\Bar{K}NNNN$,  and $\Bar{K}NNNNN$ are $25\sim28$ MeV, $45\sim50$ MeV, $68\sim76$ MeV, and $70\sim81$ MeV, respectively.
In Ref.~\cite{Kezerashvili:2021ren} the studies of the $\Bar{K}NNN$ system and $\Bar{K}\Bar{K}NN$ system in different methods were reviewed.  The binding energies of the $\Bar{K}NNN$ and $\Bar{K}\Bar{K}NN$ states are in the range of $17\sim110$ MeV and $31\sim117$ MeV. 

The $D_{s0}^{*}(2317)$ state, discovered by the BABAR Collaboration in 2003~\cite{BaBar:2003oey} and subsequently confirmed by the CLEO Collaboration~\cite{CLEO:2003ggt}, Belle Collaboration~\cite{Belle:2003guh}, and BESIII Collaboration~\cite{BESIII:2017vdm}, is a strange-charmed scalar meson and lies about 45 MeV below the $DK$ threshold. It is not easy to interpret the $D_{s0}^{*}(2317)$ as a conventional $c\Bar{s}$ state because its mass is far below the lightest $c\Bar{s}$ scalar state predicted in the conventional quark model~\cite{Godfrey:1985xj,vanBeveren:2003kd,Browder:2003fk,Barnes:2003dj,Cheng:2003kg,Chen:2004dy,Dmitrasinovic:2005gc,Zhang:2018mnm}. 
On the other hand, many studies support that it is a molecular state dynamically generated by the Weinberg-Tomozawa $DK$ interaction~\cite{Guo:2006fu,Gamermann:2006nm,Guo:2008gp,Guo:2009ct,Cleven:2010aw,MartinezTorres:2011pr,Guo:2015dha,Altenbuchinger:2013gaa,Altenbuchinger:2013vwa,Geng:2010vw,Guo:2019dpg}, which is confirmed by many lattice QCD studies~\cite{Liu:2012zya,Mohler:2013rwa,Lang:2014yfa,Bali:2017pdv,Alexandrou:2019tmk}.  

The $DK$ bound-state picture naturally inspired studies of  $DDK$, $D\bar{D}K$~\cite{Wu:2020job}, and $DKK$ three-body systems~\cite{SanchezSanchez:2017xtl,Wu:2019vsy,MartinezTorres:2018zbl,Debastiani:2017vhv}.~\footnote{See, e.g., Ref.~\cite{Wu:2022ftm} for a short review.} In Ref.~\cite{Wu:2019vsy},  the $DDK$ system forms a three-body molecule with a binding energy of about 70 MeV. In the coupled channel approach~\cite{MartinezTorres:2018zbl}, the $DDK$ bound state was found to lie about 90 MeV below the $DDK$ threshold. In Ref.~\cite{Debastiani:2017vhv}, it was found that the $KK$ repulsion is of the same magnitude as the $DK$ attractive interaction, which prevents the $DKK$ system from binding. Naively, one expects that the $DDKK$ system may bind because, with one more $D$ meson, the extra $DK$ attraction can help bind the four-body system. Indeed, the $DDDK$ system was shown to bind with a binding energy of $74\sim106$ MeV~\cite{Wu:2019vsy}.
To obtain a complete picture of multi-hadron states composed of different numbers of $D$ mesons and kaons, similar to the light nuclear system and nucleon-antikaon system, we study in this work the $DDKK$ and $D\Bar{D}K\Bar{K}$ systems with the Gaussian expansion method (GEM)~\cite{Hiyama:2003cu, Kamimura:1988zz}. 

This article is organized as follows. In Sec.\ref{sec:formalism}, we explain how to use the GEM to solve the four-body  Schr\"{o}dinger equation and construct the four-body wave functions. In Sec.\ref{sec:potential}, we explain how we determine the subsystem potentials involved in the four-body systems. We present and discuss the results in Sec.\ref{sec:result}, followed by a summary in Sec.\ref{sum}.

\section{Theoretical Formalism}
\label{sec:formalism}
\begin{figure}[!htbp]
\begin{center}
\begin{overpic}[scale=0.68]{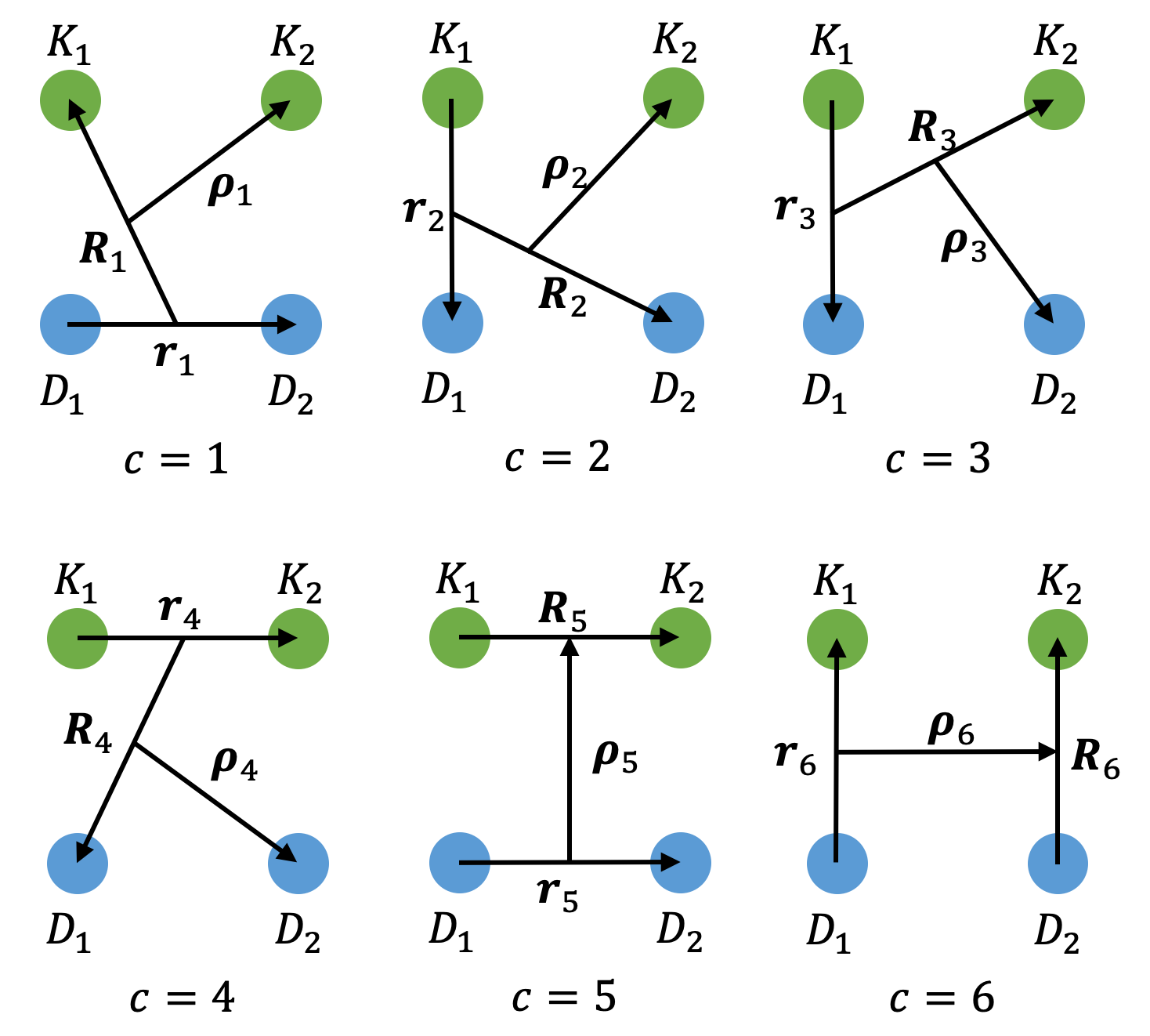}
\end{overpic}
\caption{ Six Jacobi coordinates of the $DDKK$ system. Symmetrization is implicit between $D$ mesons and between kaons }
\label{DDKK}
\end{center}
\end{figure}

To obtain the binding energies and wave functions of the four-body systems, we solve the four-body Schr\"{o}dinger equation with the GEM, which has been widely applied to investigate few-body nuclear~\cite{Hiyama:2003cu} and hadronic~\cite{Hiyama:2005cf, Yoshida:2015tia} systems. The four-body Schr\"{o}dinger equation reads
\begin{equation}
    [T+\sum_{1=i<j}^{6}V(r_{ij})-E]\Psi_{J}^{Total}=0\,,
\end{equation}
where $T$ is the kinetic-energy operator, $V(r_{ij})$ are the two-body potentials between particle $i$ and particle $j$, and $J$ is the total angular momentum of the four-body system. 
The four-body total wave function $\Psi_{J}^{Total}$  is expressed as a sum of the wave functions of the eighteen rearrangement coordinates~\cite{Hiyama:2003cu}:
\begin{equation}
    \Psi_{J}^{Total}=\sum_{c,\alpha}C_{c,\alpha}\Phi_{J,\alpha}^{c}(\pmb{r}_{c},\pmb{R}_{c},\pmb{\rho}_{c})\quad (c=1-18)\,,
\end{equation}
where $C_{c,\alpha}$ are the expansion coefficients of relevant bases,  and $c=1 - 18$ denote the eighteen Jacobi coordinates of the four-body system. Thanks to the symmetry of the $DDKK$ system, the coordinate number reduces to six, as shown in Fig.~\ref{DDKK}. $\alpha$ is a collection of parameters $\{l, L, \lambda, L', \Lambda, t, T, I\}$. Here, $l$, $L$, and $\lambda$ are the orbital angular momenta of the coordinates $r$, $R$, and $\rho$ in each channel. $L'$ is the coupling of $l$ and $L$, the total orbital angular momentum $\Lambda$ is the coupling of $L'$ and $\lambda$. $t$ and $T$ are the isospin of the coordinates $r$ and $R$ in each channel, and the total isospin $I$ is the coupling of $T$ and $t$ (or the last particle), which will be discussed later.

Because the spin of $D$  and $K(\bar{K})$ is zero, there is no need to consider the spin wave function.  Therefore, the wave function of each channel can be written as 
\begin{equation}
   \begin{aligned}
    \Phi_{J,\alpha}^{c}(\pmb{r}_{c},\pmb{R}_{c},\pmb{\rho}_{c})=&\left[\left[\phi^{G}_{n_{c}l_{c}}(\pmb{r}_{c})\psi^{G}_{N_{c}L_{c}}(\pmb{R}_{c})\right]_{L'_{c}} \chi^{G}_{\nu_{c}\lambda_{c}}(\pmb{\rho}_{c}) \right]_{\Lambda}\\
    & \otimes H_{tT,I}^{c}\,,
   \end{aligned}
\end{equation}
where $\phi^{G}_{n_{c}l_{c}}$, $ \psi^{G}_{N_{c}L_{c}}$, and $ \chi^{G}_{\nu_{c}\lambda_{c}}$ are the spatial wave functions and $H_{tT,I}^{c} $  is the total isospin wave function. Three spatial wave functions can be expanded in terms of Gaussian functions of the Jacobi coordinates $\pmb{r}$, $\pmb{R}$, and $\pmb{\rho}$:
\begin{equation}
    \begin{aligned}
    &\phi^{G}_{nlm}(\pmb{r})=\phi^{G}_{nl}(r)Y_{lm}(\hat{\pmb{r}})\,,~ \phi^{G}_{nl}(r)=N_{nl} r^{l} e^{-\nu_{n}r^{2}}\,, \\
    &\psi^{G}_{NLM}(\pmb{R})=\psi^{G}_{NL}(r)Y_{LM}(\hat{\pmb{R}})\,,~ \psi^{G}_{NL}(R)=N_{NL} R^{L} e^{-\lambda_{N}R^{2}}\,,\\
    &\chi^{G}_{\nu\lambda\mu}(\pmb{\rho})=\chi^{G}_{\nu\lambda}(\rho)Y_{\lambda\mu}(\hat{\pmb{\rho}})\,,~ \chi^{G}_{\nu\lambda}(\rho)=N_{\nu\lambda} \rho^{\lambda} e^{-\omega_{\nu}\rho^{2}}\,,
    \end{aligned}
\end{equation}
where  $N_{nl}$, $N_{NL}$, and $N_{\nu\lambda}$  are the  normalization constants, and the relevant parameters $\nu_{n}$, $\lambda_{N}$, and $\omega_{\nu}$ are given by
\begin{equation}
   \begin{aligned}
    &\nu_{n}=1/r^{2}_{n},~~ r_{n}=r_{1}a^{n-1},~~~~~~~~~~(n=1, 2, ..., n_{max})\\
    &\lambda_{N}=1/R^{2}_{N},~~ R_{N}=R_{1}A^{N-1},~~(N=1, 2, ..., N_{max})\\
    &\omega_{\nu}=1/\rho^{2}_{\nu},~~ \rho_{\nu}=\rho_{1}a'{}^{\nu-1},~~~~~~~~~(\nu=1, 2, ..., \nu_{max})
   \end{aligned}
\end{equation}
where $\{n_{max}, r_{1}, a~ \mbox{or}~$$r_{max} \}$, $\{N_{max}, R_{1},  A~ \mbox{or}~ $ $ R_{max} \}$, and $\{\nu_{max}, \rho_{1},  a'~ \mbox{or}~ $ $ \rho_{max} \}$ are Gaussian basis parameters given in Table~\ref{quantum-number}.

As shown in Fig.~\ref{DDKK}, there are two types of configurations, i.e., K-type ($c=1-4$) and H-type ($c=5,6$). As a result, the total isospin wave function $H_{tT,I}^{c}$ has two coupling ways, $H_{tT,I}^{K}$ and $H_{tT,I}^{H}$, which can be written as 
\begin{equation}
   \begin{aligned}
    &H_{tT,I}^{K}=[[[\eta_{\frac{1}{2}}(i)\eta_{\frac{1}{2}}(j)]_{t}\eta_{\frac{1}{2}}(k)]_{T}\eta_{\frac{1}{2}}(n)]_{I} \,,\\
    &H_{tT,I}^{H}=[[\eta_{\frac{1}{2}}(i)\eta_{\frac{1}{2}}(j)]_{t}[\eta_{\frac{1}{2}}(k)\eta_{\frac{1}{2}}(n)]_{T}]_{I} \,,
   \end{aligned}
\end{equation}
where $\eta_{\frac{1}{2}}(i)$ is the isospin wave function of particle $i$. The possible values of $t$ and $T$ are listed in Table~\ref{quantum-number}

\begin{table}[!htbp]
    \centering
    \caption{Four-body isospin space and the Gaussian range parameters for the $I(J^{P}) = 0(0^{+})$ configuration of the $DDKK$ system. Lengths are in units of fm.}
    \begin{tabular}{ccccccccccccc}
    \hline\hline
     $c$ &  $t$& $T$ &$n_{max}$ & $r_{1}$ & $r_{max}$ &$N_{max}$ & $R_{1}$ & $R_{max}$&$\nu_{max}$ & $\rho_{1}$ & $\rho_{max}$  \\\hline
     1 & 1 & 1/2 & 8 & 0.2 & 10 & 6 &  0.4 &10 & 6 &  0.6 &10  \\
     2 & 0(1) & 1/2 & 8 & 0.1 & 10 & 5 &  0.3 &10 & 6 &  0.6 &10 \\
     3 & 0(1) & 1/2 & 8 & 0.3 & 10 & 5 &  0.6 &10 & 6 &  0.6 &10 \\
    4 & 1 & 1/2 & 8 & 0.1 & 10 & 5 & 0.4 &10 &  6 &  0.6 &10  \\
      5 & 1 & 1 & 8 & 0.2 & 9 & 6 & 0.1 &10 &  6 &  0.2 &8 \\
     6 & 0(1) & 0(1) & 8 & 0.1 & 8 & 5 & 0.1 &10 &  8 &  0.2 &8  \\
    \hline\hline
    \end{tabular}
    \label{quantum-number}
\end{table}

\section{Two-body potentials}
\label{sec:potential}

\subsection{The $DK$ interaction}
In this work, we employ the $DK$ interaction used in Refs.~\cite{MartinezTorres:2018zbl, Wu:2019vsy}. The dominant contribution to the $S$-wave $DK$ interaction is the Weinberg-Tomozawa(WT) term between a $D$ meson and a kaon, which can be formulated as a standard quantum  mechanical potential in the non-relativistic limit,
\begin{equation}
    V_{DK}(q)=-\frac{C_{W}(I)}{2f_{\pi}^{2}}\,,
    \label{DKq}
\end{equation}
where the pion decay constant $f_{\pi}\approx$  130 MeV and the strength of the WT term $C_{W}(I)$ depends on the isospin of the $DK$ system,
\begin{equation}
    C_{W}(0)=2,\quad C_{W}(1)=0 \,.
\end{equation}
After Fourier transform and with a local Gaussian regulator, the $DK$ potential in coordinate space reads~\cite{Wu:2019vsy}
\begin{equation}
    V_{DK}(r,R_{C})=-\frac{C_{W}(I)}{2f_{\pi}^{2}}\frac{e^{-(r/R_{C})^2}}{\pi^{3/2}R_{C}^{3}} \,.
\end{equation}

According to chiral perturbation theory (ChPT)~\cite{Altenbuchinger:2013vwa}, the leading-order $S$-wave $DK$ interaction in isospin zero is attractive, while the next-to-leading (NLO) order correction is weakly repulsive. The LO $DK$ interaction reads,
\begin{equation}
    V_{DK}(r,R_{C})=C(R_{C})\frac{e^{-(r/R_{C})^2}}{\pi^{3/2}R_{C}^{3}}\,,
\end{equation}
where $R_{C}$ is the cutoff, and $C(R_{C})$ is the running coupling constant describing the strength of the LO $DK$ interaction. We should also add a short-range repulsive core in the $DK$ interaction as the NLO correction so the total $DK$ interaction can be written as 
\begin{eqnarray}
    V_{DK}(r, R_{C}) &=& C_{S}\frac{e^{-(r/R_{S})^2}}{\pi^{3/2}R_{S}^{3}}+ C(R_{C})\frac{e^{-(r/R_{C})^2}}{\pi^{3/2}R_{C}^{3}}\nonumber\\
    &=&C'_{S}e^{-(r/R_{S})^2}+C'_{L}e^{-(r/R_{C})^2}\,,
\end{eqnarray}
where $C'_{S}$, $C'_{L}$ are coupling constants, and $R_{S}$, $R_{C}$ are cutoffs for the repulsive and attractive potentials. The uncertainty coming from the NLO corrections can be estimated by varying the cutoff $R_{C}$ within a sensible range. Specifically, we require that $C'_{S} > |C'_{L}|$ and $R_{S} < R_{C}$ and take $R_{S}=0.5$ fm, $R_{c} = 1.0, 2.0, 3.0$ fm, $C'_{S}=0, 500, 1000, 3000$ MeV. For each set of these parameter values, $C'_{L}$ is determined by reproducing the $D^*_{s0}(2317)$ as a $DK$ bound state with a binding energy of 45 MeV.

\subsection{The $DD$ interaction}
Due to a lack of data, we turn to the one-boson exchange (OBE) model~\cite{Machleidt:1987hj,Machleidt:1989tm} to describe the $DD$ interaction. In the OBE model, the potential between two hadrons is generated by exchanging light mesons (i.e., $\pi$, $\sigma$, $\rho$, and $\omega$). The OBE model has been successfully employed in accurately describing the nuclear force~\cite{Machleidt:1987hj,Machleidt:1989tm} and predicting the existence of heavy hadron molecules~\cite{Voloshin:1976ap}. In recent years, it has been widely applied to study newly discovered hadronic molecules~\cite{Liu:2019stu,Liu:2021pdu,Luo:2021ggs,Wu:2021kbu,Liu:2019zvb,Wu:2021dwy,Pan:2020xek,Pan:2022xxz,Sun:2011uh,Chen:2019asm,Li:2012ss,He:2015mja,Lee:2011rka}.

The $DD$ potential is generated by exchanging $\sigma$, $\rho$, $\omega$~\cite{Liu:2019stu}:
\begin{equation}
    V_{DD}(r, \Lambda) = V_{\sigma}(r, \Lambda)+V_{\rho}(r, \Lambda)+V_{\omega}(r, \Lambda)\,,
\end{equation}
where a form factor and a cutoff $\Lambda$ regularize the contribution of each light meson. The particular contribution of each meson is  written as~\cite{Liu:2019stu}
\begin{eqnarray}
    V_{\sigma}(r,\Lambda)&=& -g_{\sigma}^{2}m_{\sigma}W_{C}\left(m_{\sigma} r, \frac{\Lambda}{m_{\sigma}} \right)\,,\\
    V_{\rho}(r,\Lambda)&=& +\Vec{\tau}_{1}\cdot\Vec{\tau}_{2} g_{\rho}^{2}m_{\rho}W_{C}\left(m_{\rho} r, \frac{\Lambda}{m_{\rho}} \right)\,,\\
    V_{\omega}(r,\Lambda)&=& +g_{\omega}^{2}m_{\omega}W_{C}\left(m_{\omega} r, \frac{\Lambda}{m_{\omega}} \right)\,,
\end{eqnarray}
where
\begin{equation}
    W_{C}(x,\lambda) = \frac{e^{-x}}{4\pi x}-\lambda\frac{e^{-\lambda x}}{4\pi\lambda x}-\frac{(\lambda^2-1)}{2\lambda}\frac{e^{-x}}{4\pi}\,.
\end{equation}

The masses of the exchanged bosons are $m_{\sigma}=0.6$ GeV, $m_{\rho}=0.77$ GeV, $m_{\omega}=0.78$ GeV, and the couplings are  $g_{\rho}=g_{\omega}=2.6$, $g_{\sigma}=3.4$. The cutoff $\Lambda$ is determined by reproducing
the mass of the $X(3872)$ as a $D\bar{D}^*$ molecule, yielding $\Lambda = 1.01^{+0.19}_{-0.10}$ GeV~\cite{Liu:2019stu}. In this work, we set the cutoff to $\Lambda = 1.0$ GeV and neglect the small uncertainty due to the exploratory nature of the present work.

\subsection{The $KK$ interaction}

The kaon-kaon interaction has been widely investigated in studies of $a_{0}(980)$~\cite{Weinstein:1990gu, Dudek:2016cru}, $KK\Bar{K}$~\cite{MartinezTorres:2011gjk}, $K\Bar{K}N$~\cite{Jido:2008kp}, $\Bar{K}\Bar{K}N$~\cite{Kanada-Enyo:2008wsu}, etc. Considering that the typical kinetic energy of the kaon in such systems is small compared to its mass, we adopt the non-relativistic $KK(K\Bar{K})$ potential in the present work. For the $S$-wave $KK$ effective potential, because of the Bose-Einstein statistics, $I=0$ is forbidden. Therefore, we only consider the $I=1$ interaction. Following Refs.~\cite{MartinezTorres:2011gjk, Kanada-Enyo:2008wsu, Jido:2008kp}, the $KK$ and $K\Bar{K}$ potentials can be written as one Gaussian function:
\begin{equation}
    V_{KK}(r) = v_{0} e^{-(r/b)^2}\,,
\end{equation}
where $v_{0}$ and $b$ are the strength of the potential and the interaction range, respectively. The interaction ranges of $KK$ and $K\Bar{K}$ are assumed to be the same. 
The strength of the $KK$ interaction is determined by reproducing the lattice QCD scattering length $a_{K^{+}K^{+}}=-0.14$~\cite{Beane:2007uh}. The strength of the $K\Bar{K}$ interaction is determined by fitting to the masses and widths of $f_{0}(980)$ and $a_{0}(980)$~\cite{ParticleDataGroup:2022pth} assuming that $f_{0}(980)$ and $a_{0}(980)$ are quasibound states of $K\Bar{K}$ in the  $I=0$ and $I=1$ channels, respectively. Here, We adopt the parameters for the two cases studied in Refs.~\cite{MartinezTorres:2011gjk, Kanada-Enyo:2008wsu, Jido:2008kp}. In Case A, $v^{KK}_{0}=104$ MeV,$v_{0}^{K\Bar{K}}=-630-210i$ MeV, and $b=0.66$ fm, while in Case $B$,  $v^{KK}_{0}=313$ MeV, $v_{0}^{K\Bar{K}}=-1155-283i$ MeV, and  $b=0.47$ fm.

\begin{table*}[!htbp]    
    \centering
    \caption{Binding energies, expectation values of the Hamiltonian (potential and kinetic energies),  and rms radii of the four-body system $DDKK$. Energies are in units of MeV and radii are in units of fm. The numbers outside and inside the brackets represent cases A and B.  The cutoffs $R_{S}$ and $R_{C}$ are in units of fm, and the coupling constants $C'_{S}$ and $C'_{L}$ are in units of MeV. \label{results1} }
    \scalebox{0.965}{
    \begin{tabular}{cccccccccccc}
    \hline\hline
          $R_{S}$&$R_{C}$&$C_{S}^{'}$ & $C_{L}^{'}$ & $E$ & $\left \langle T \right \rangle $ & $\left \langle V_{DD}\right \rangle$ & $\left \langle V_{DK}\right \rangle$ & $\left \langle V_{KK}\right \rangle$& $r_{DD}$ & $r_{DK}$ & $r_{KK}$\\ \hline
         \multirow{4}{*}{0.5}&\multirow{4}{*}{1}
            &0 & $-$320.1 & $-$154.78($-$151.84) & 239.38(233.70) & $-$9.07($-$8.89) & $-$396.42($-$389.32) & 11.33(12.67) & 1.03(1.04) & 1.20(1.22) & 1.54(1.57) \\
         &&500 & $-$455.4 & $-$152.03($-$150.40) & 189.39(187.46) & $-$7.58($-$7.52) & $-$342.14($-$339.24) & 8.30(8.90) & 1.14(1.14) & 1.33(1.34) & 1.70(1.72) \\
         &&1000 & $-$562.6 & $-$150.20($-$149.09) & 174.04(173.10) & $-$6.85($-$6.83) & $-$324.12($-$322.40) & 6.72(7.04) & 1.20(1.21) & 1.41(1.42) & 1.81(1.83) \\
         &&3000 & $-$838.7 & $-$146.11($-$145.43) & 181.27(180.73) & $-$6.24($-$6.24) & $-$325.73($-$324.63) & 4.59(4.91) & 1.31(1.31) & 1.58(1.59) & 2.04(2.05) \\ \hline
         \multirow{4}{*}{0.5}&\multirow{4}{*}{2}
           &0 & $-$149.1 & $-$145.80($-$145.26) & 113.17(112.63) &  $-$4.95($-$4.92) & $-$258.35($-$257.35) & 4.32(4.37) & 1.43(1.44) & 1.69(1.70) & 2.15( 2.17) \\
         &&500 & $-$178.4 & $-$143.92($-$143.64) & 95.39(95.33) &  $-$4.07($-$4.07) & $-$238.19($-$237.83) & 2.96(2.93) & 1.58(1.58) & 1.87(1.88) & 2.39(2.40) \\
         &&1000 & $-$195.0 & $-$142.79($-$142.57) & 95.12(95.09) & $-$3.84($-$3.85) & $-$236.55($-$236.26) & 2.48(2.45) & 1.64(1.64) & 1.97(1.97) & 2.52(2.53) \\
         &&3000 & $-$225.9 & $-$140.59($-$140.41) & 102.70(102.65) & $-$3.80($-$3.80) & $-$241.43($-$241.17) & 1.94(1.91) & 1.70(1.70) & 2.12(2.12) &  2.73(2.74) \\ \hline
         \multirow{4}{*}{0.5}&\multirow{4}{*}{3}
            &0 & $-$107.0 & $-$142.36($-$142.17)& 73.80(73.71) & $-$3.30($-$3.29) & $-$215.21($-$214.89) & 2.34(2.29) & 1.75(1.75) & 2.07(2.08) & 2.64(2.65)\\
         &&500 & $-$119.4 & $-$141.13($-$141.03) & 64.78(64.81) & $-$2.76($-$2.76) & $-$204.80($-$204.67) & 1.65(1.59) & 1.90(1.90) & 2.28(2.28) & 2.90(2.90) \\
         &&1000 & $-$125.6 & $-$140.38($-$140.29) & 65.27(65.29) & $-$2.65($-$2.65) &  $-$204.43($-$204.31) & 1.43(1.38) & 1.96(1.96) & 2.37(2.37) & 3.02(3.03) \\
         &&3000 & $-$136.2 & $-$138.95($-$138.87) & 69.59(69.61) & $-$2.64($-$2.64) & $-$207.07($-$206.96) & 1.16(1.12) & 2.02(2.02) & 2.52(2.52) & 3.23(3.24) \\ \hline\hline
    \end{tabular}}\\
\end{table*}

\section{Results and Discussions}
\label{sec:result}

\begin{figure}[!htbp]
\begin{center}
\begin{overpic}[scale=0.4]{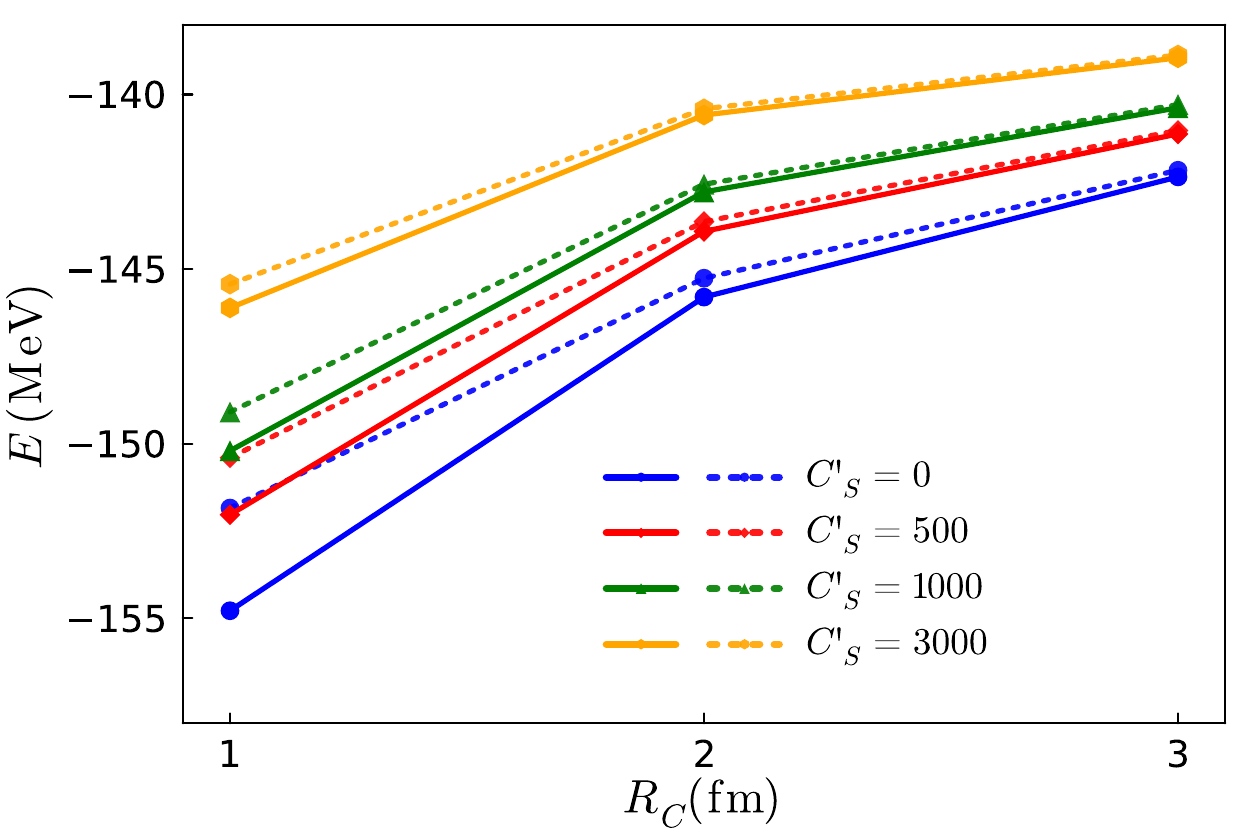}
\end{overpic}
\caption{ Binding energies of the $DDKK$ system as functions of the cutoff $R_{C}$. The solid lines and dashed lines correspond to Case $A$ and Case $B$, respectively. Blue, red, green, and orange lines are for $C'_{S}=0$, $500$, $1000$, $3000$ MeV, respectively. }
\label{BE}
\end{center}
\end{figure}

\begin{figure*}[!htbp]
\centering
\begin{minipage}[t]{0.33\textwidth}
\centering
\includegraphics[width=\textwidth]{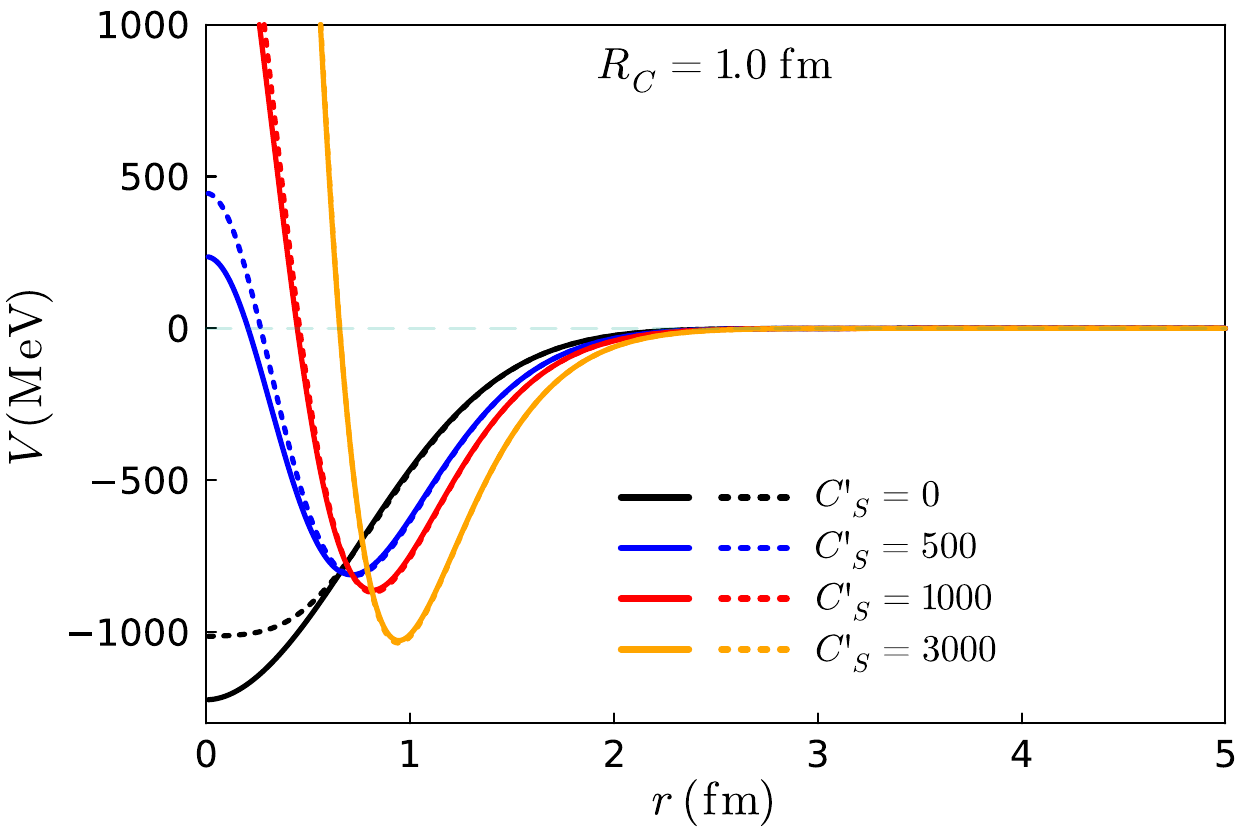}
%\subcaption{1}
\end{minipage}
\begin{minipage}[t]{0.33\textwidth}
\centering
\includegraphics[width=\textwidth]{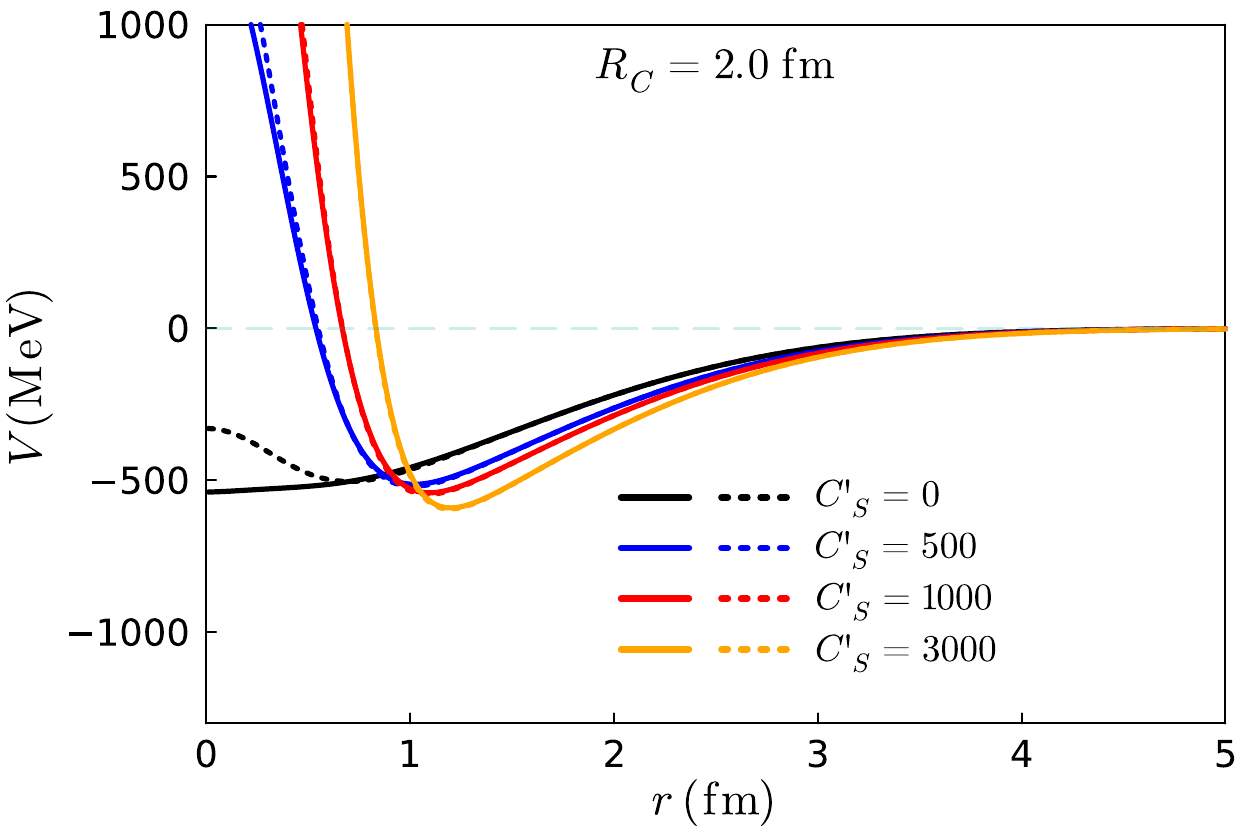}
%\subcaption{2}
\end{minipage}
\begin{minipage}[t]{0.33\textwidth}
\centering
\includegraphics[width=\textwidth]{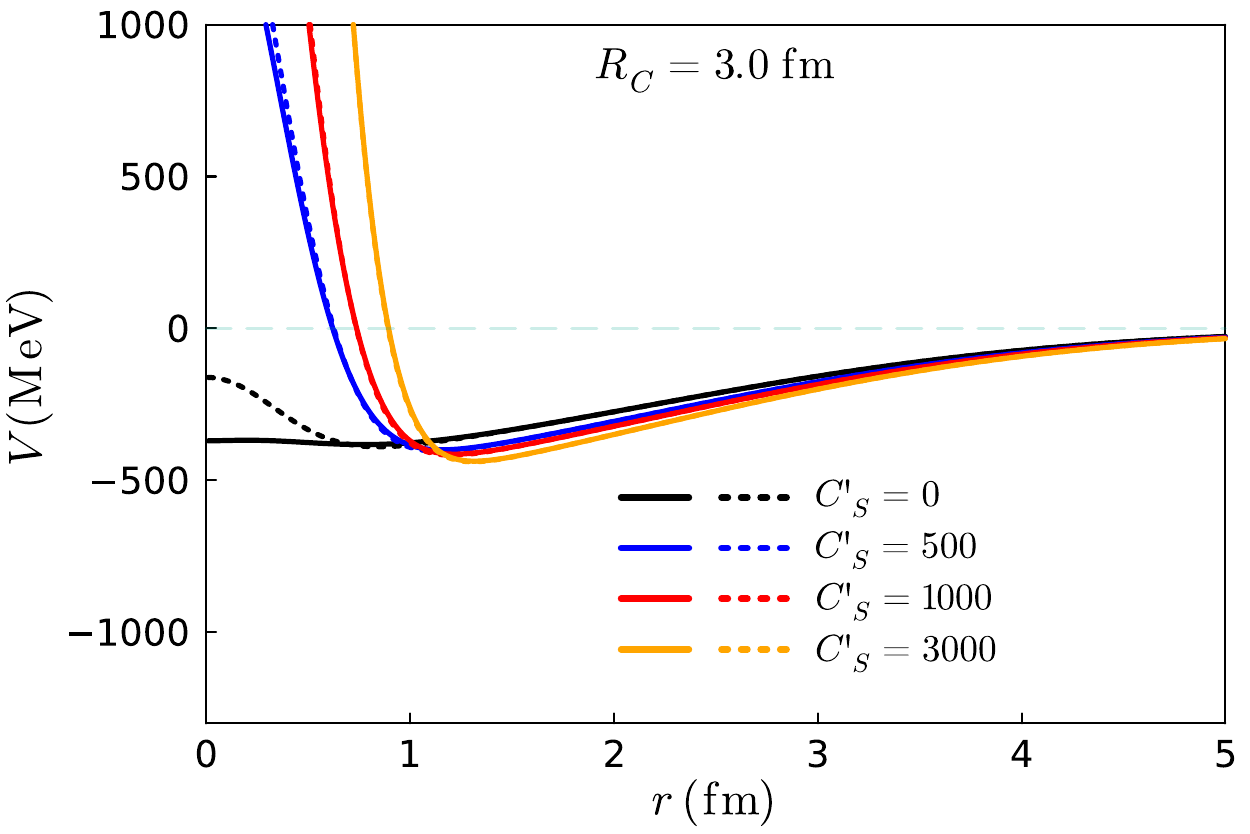}
%\subcaption{2}
\end{minipage}
\caption{Total potential of the $DDKK$ system, the solid and dashed lines are for Case $A$ and Case $B$. The black, blue, red, and orange lines are for $C'_{S}=0,500,1000,3000$ MeV. The left, middle, and right figures are the total potential obtained for $R_{C}=1.0,2.0,3.0$ fm.}
\label{Vtotal}
\end{figure*}

The binding energies, expectation values, and root-mean-square (rms) radii of the $DDKK$ system are given in Table~\ref{DDKK}. The binding energies as functions of potential parameters are shown in Fig.~\ref{BE}. Clearly, for all parameters studied, the $DDKK$ system is always bound with a binding energy of $138\sim 155$ MeV. As the strength of the repulsive core $C'_{S}$ and the cutoff $R_{C}$ increases, the binding energy decreases. In addition, as  $R_{C}$ grows, the differences among different $C'_{S}$ and between Case $A$ and Case $B$ become smaller. This trend can also be seen in the potentials for different $R_{C}$, as shown in Fig.~\ref{Vtotal}. As $R_{C}$ increases, the total potential becomes flatter, and the differences between Case $A$ and Case $B$ and among different $C'_{S}$ decrease, especially for the range of our interest, $1$ fm $< r < 2$ fm, which is responsible for the trend observed above. From the expectation values of the potentials, one concludes that the $DK$ interaction plays a dominant role. The strength of the repulsive $KK$ interaction is compatible with the strength of the attractive $DD$ interaction, which is much smaller than the strength of the attractive $DK$ interaction. Therefore, the differences between Case $A$ and Case $B$ are minor.
The kinetic energy of the four-body system is much smaller than the kaon mass, which justifies the use of a non-relativistic potential for the kaon-kaon interaction. The rms radius of the $KK$ subsystem is larger than those of the $DK$ subsystem and the $DD$ subsystem. As  $R_{C}$ and $C'_{S}$ increase, so do the rms radii, and the rms radii of Case $B$ are a bit larger than those of Case $A$. The rms radii of each subsystem are all in the range of $1.0 \sim 3.3$ fm, consistent with the typical size of a hadronic molecule.

In Ref.~\cite{Wu:2019vsy}, the $DDDK$ system is shown to have a binding energy of $91 \sim 107$ MeV, which is about 50 MeV smaller than the binding energy of the $DDKK$ system. In both systems, the $DK$ interactions are dominant. Intuitively, there are four $DK$ pairs in the $DDKK$ system, one more than in the $DDDK$ system, which can qualitatively explain the difference of binding energy.

\begin{table*}[!htbp]    
    \setlength{\tabcolsep}{5.5pt}
    \centering
    \caption{Binding energies, expectation values of the Hamiltonian (potential and kinetic energies) and rms radii of the four-body system $D\Bar{D}K\Bar{K}$. Energies are in units of MeV and radii are in units of fm. The relevant parameter values are $R_{S}=$ 0.5 fm and $C'_{S}=$ 0 MeV.
    \label{DDbKKb} }
    \begin{tabular}{cccccccccccc}
    \hline\hline
          $R_{C}$& $C_{L}^{'}$ & $E$ & $\left \langle T \right \rangle $ & $\left \langle V_{D\Bar{D}}\right \rangle$ & $\left \langle V_{DK+\Bar{D}\Bar{K}}\right \rangle$ & $\left \langle V_{\Bar{D}K+D\Bar{K}}\right \rangle$& $\left \langle V_{K\Bar{K}} \right \rangle$&  $r_{D\Bar{D}}$ & $r_{DK/\Bar{D}\Bar{K}}$ & $r_{\Bar{D}K/D\Bar{K}}$& $r_{K\Bar{K}}$ \\ \hline
         \multicolumn{12}{c}{$b=$ 0.66 fm ~~ $v_{0}^{K\Bar{K}}=$ $-$630$-$210$i$ MeV}\\\hline
         1& $-$320.1 & $-$156.15$-$53.35$i$ & 307.65 & $-$12.68 & $-$228.98 & $-$62.10 & $-$160.04  & 1.07 & 1.08 & 1.16 & 1.14  \\
         2& $-$149.1 & $-$133.44$-$42.01$i$ & 186.93 & $-$7.09 & $-$142.76 & $-$44.49 & $-$126.04 & 1.46 & 1.51 & 1.56 & 1.40  \\
         3& $-$107.1 & $-$123.60$-$37.42$i$ & 147.15 & $-$4.79 & $-$116.33 & $-$37.39 & $-$112.25  & 1.81 & 1.85 & 1.89 & 1.57  \\ \hline
         \multicolumn{12}{c}{$b=$ 0.47 fm ~~ $v^{K\Bar{K}}_{0} =$ $-$1155$-$283$i$ MeV}\\\hline
         1& $-$320.1 & $-$162.95$-$53.17$i$ &  368.87 & $-$13.32 & $-$234.25 & $-$67.22  & $-$217.01 & 1.04 & 1.04 & 1.10 & 1.04  \\
         2& $-$149.1 & $-$135.33$-$41.60$i$ & 231.47 & $-$7.24 & $-$144.20 & $-$45.56 & $-$169.80  & 1.44 & 1.48 & 1.53 & 1.30   \\
         3& $-$107.1 & $-$124.37$-$37.04$i$ & 186.49 & $-$4.86 & $-$116.99 & $-$37.83 & $-$151.18 & 1.80 & 1.83 & 1.87 & 1.48  \\ \hline\hline
    \end{tabular}
\end{table*}

\begin{figure*}[!htbp]
\centering
\begin{minipage}[t]{0.33\textwidth}
\centering
\includegraphics[width=\textwidth]{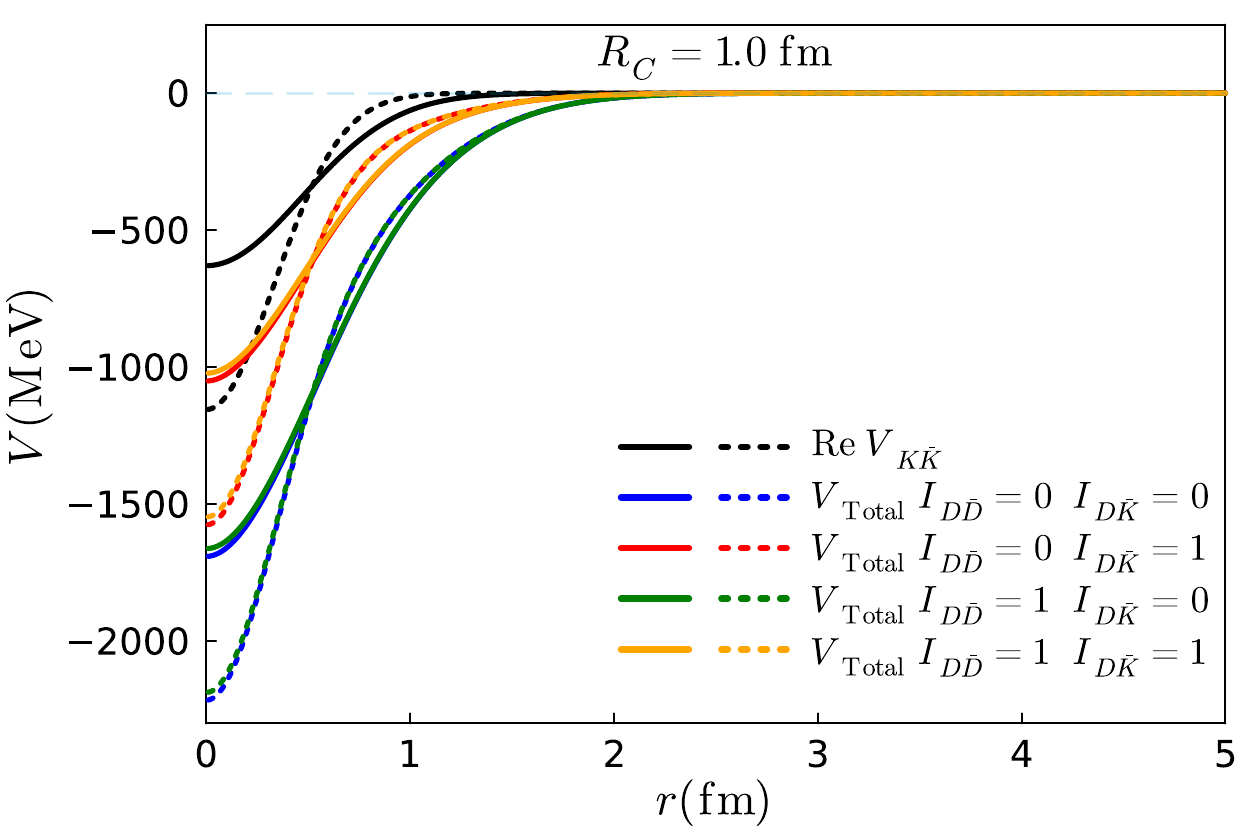}
%\subcaption{1}
\end{minipage}
\begin{minipage}[t]{0.33\textwidth}
\centering
\includegraphics[width=\textwidth]{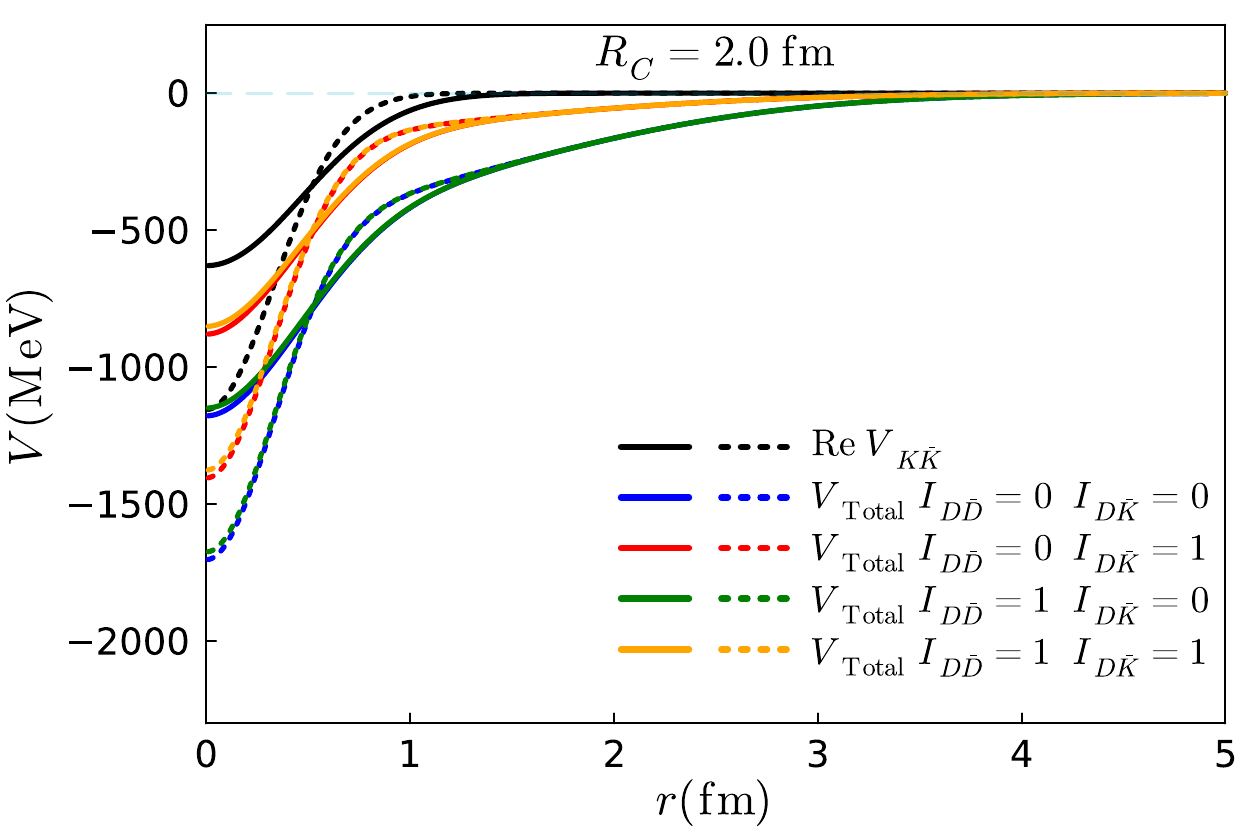}
%\subcaption{2}
\end{minipage}
\begin{minipage}[t]{0.33\textwidth}
\centering
\includegraphics[width=\textwidth]{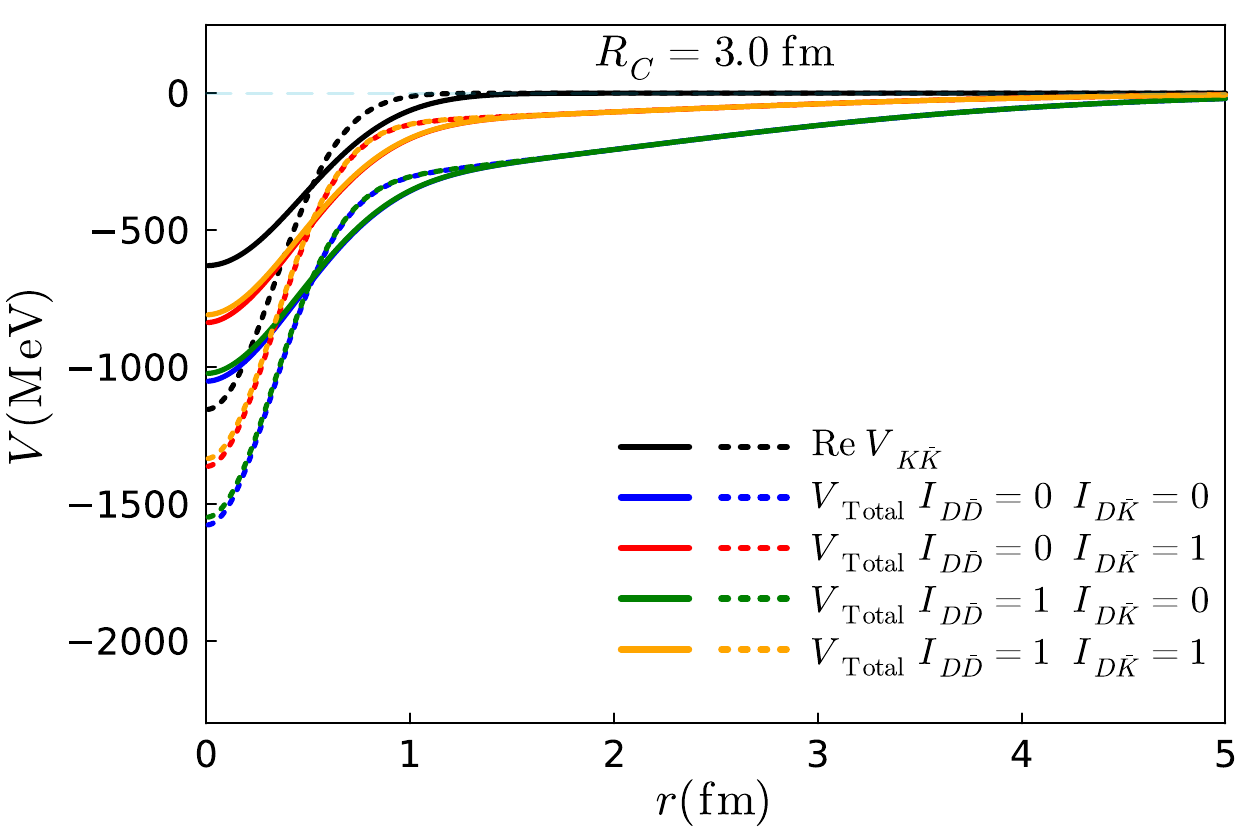}
%\subcaption{2}
\end{minipage}
\caption{Total potential of the $D\Bar{D}K\Bar{K}$ system, the solid and dashed lines are for Case $A$ and Case $B$. The black line is for the real part of the $K\Bar{K}$ interaction. The left, middle, and right figures are the total potential obtained for $R_{C}=1.0,2.0,3.0$ fm.}
\label{DD-DK-KK}
\end{figure*}

Next, we investigate the $D\Bar{D}K\Bar{K}$ system. There are no identical particles in this system. Therefore, all the 18 Jacobian channels are distinguishable. Compared to the $DDKK$ system, the $D\Bar{D}K\Bar{K}$ system is unique because the $K\Bar{K}$ interaction is complex. To study the  $D\Bar{D}K\Bar{K}$ system, we follow the method of Ref.~\cite{Jido:2008kp} in which the imaginary part of the $K\Bar{K}$ interaction is treated as a perturbative correction in the $K\Bar{K}N$ molecular state. 
They also discussed the difference between the complex energy obtained in perturbative and nonperturbative treatments for the $K\Bar{K}$ two-body system. The conclusion is that the imaginary energies obtained in both methods are similar, while the real energies increase slightly because of the higher-order corrections of the perturbative expansion~\cite{Jido:2008kp}.

We first consider the real part of the Hamiltonian in the GEM calculation. The total wave function $\Psi$ and the binding energy $E_{\rm{Re}}$ are obtained using the lowest-energy solution. The imaginary part of the energy $E_{\rm{Im}}$ is estimated by calculating the expectation value of the imaginary part of the Hamiltonian ($\rm{Im}V_{K\Bar{K}}$) with the obtained total wave function $\Psi$:
\begin{equation}
    E_{\rm{Im}}=\langle\Psi|\rm{Im}V_{K\Bar{K}}|\Psi\rangle\,.
\end{equation}
The complex energy is given as $E=E_{\rm{Re}}+iE_{\rm{Im}}$ and  the decay width is  $\Gamma=-2iE_{\rm{Im}}$.

According to chiral perturbation theory~\cite{Altenbuchinger:2013vwa}, the leading order $D\Bar{K}$ interaction is only half of the leading order $DK$ interaction in isospin zero, while in isospin one, $V_{D\Bar{K}}^{I=1}=-V_{D\Bar{K}}^{I=0}$.  It is worthy noting that in momentum space, the $I=0$ $D\Bar{K}$ can not form a bound state~\cite{Altenbuchinger:2013vwa}. However, if we simply reduce the $DK$ interactions given in Table \ref{results1} by half, they can generate a bound state.  To be consistent with the results in momentum space~\cite{Altenbuchinger:2013vwa}, we find that without the repulsive core and taking $R_{C}<0.78$ fm, one can simultaneously obtain a $DK$ bound state with a binding energy of 45 MeV and an unbound $D\Bar{K}$ system. Therefore, in the study of the $D\bar{D}K\bar{K}$ system, we take $R_{C}=0.77$ fm and $b=0.66$ fm, and obtain a bound $D\bar{D}K\bar{K}$ system with a binding energy of 166.09 MeV and a width of 117.22 MeV. We note that the smaller the $R_{C}$, the more bound the $D\bar{D}K\bar{K}$ system becomes. Therefore, to estimate the uncertainty originating from $R_{C}$ and compare this system with the $DDKK$ system, the parameters used in the $D\bar{D}K\bar{K}$ system are taken to be the same as those for the $DDKK$ system, i.e., $R_{C}=1,2,3$ fm and $b=0.47, 0.66$ fm. In addition,  we note that the short-range repulsion plays a minor role and therefore fix $C_{S}'=0$.

The results are listed in Table~\ref{DDbKKb}. The parameters of the $K\Bar{K}$ interaction have tiny effects on the $D\Bar{D}K\Bar{K}$ system. The binding energies of the $D\Bar{D}K\Bar{K}$ system range from 123 MeV to 163 MeV, compatible with those of the $DDKK$ system. Although the interaction between $D$ and kaon (antikaon) of the $D\Bar{D}K\Bar{K}$ system is weaker than that of the $DDKK$ system, the $K\Bar{K}$ interaction is strong enough to yield a total potential for the $D\Bar{D}K\Bar{K}$ system compatible with that for the $DDKK$ system as shown in Fig.~\ref{DD-DK-KK}. From the perspective of expectation values, the dominant interactions in the $D\Bar{D}K\Bar{K}$ system are the $DK$ interaction and $K\Bar{K}$ interaction. As $R_{C}$ increases and $b$ decreases, the $K\Bar{K}$ interaction plays a more important role, which can also be seen in Fig. \ref{DD-DK-KK}. The rms radii of each subsystem are all in the range of $1.0 \sim 2.0$ fm.
The imaginary part of the expectation values of the total potential $|\langle\Psi|\rm{Im}V|\Psi\rangle|$ is about $-40$ MeV, and the real part $|\langle\Psi|\rm{Re}V|\Psi\rangle|$ is about $-320$ MeV, $|\langle\Psi|\rm{Im}V|\Psi\rangle|\ll |\langle\Psi|\rm{Re}V|\Psi\rangle|$, which justifies the perturbative treatment.

\section{Summary and conclusion}
\label{sum}

\begin{figure}[!htbp]
\begin{center}
\begin{overpic}[scale=0.49]{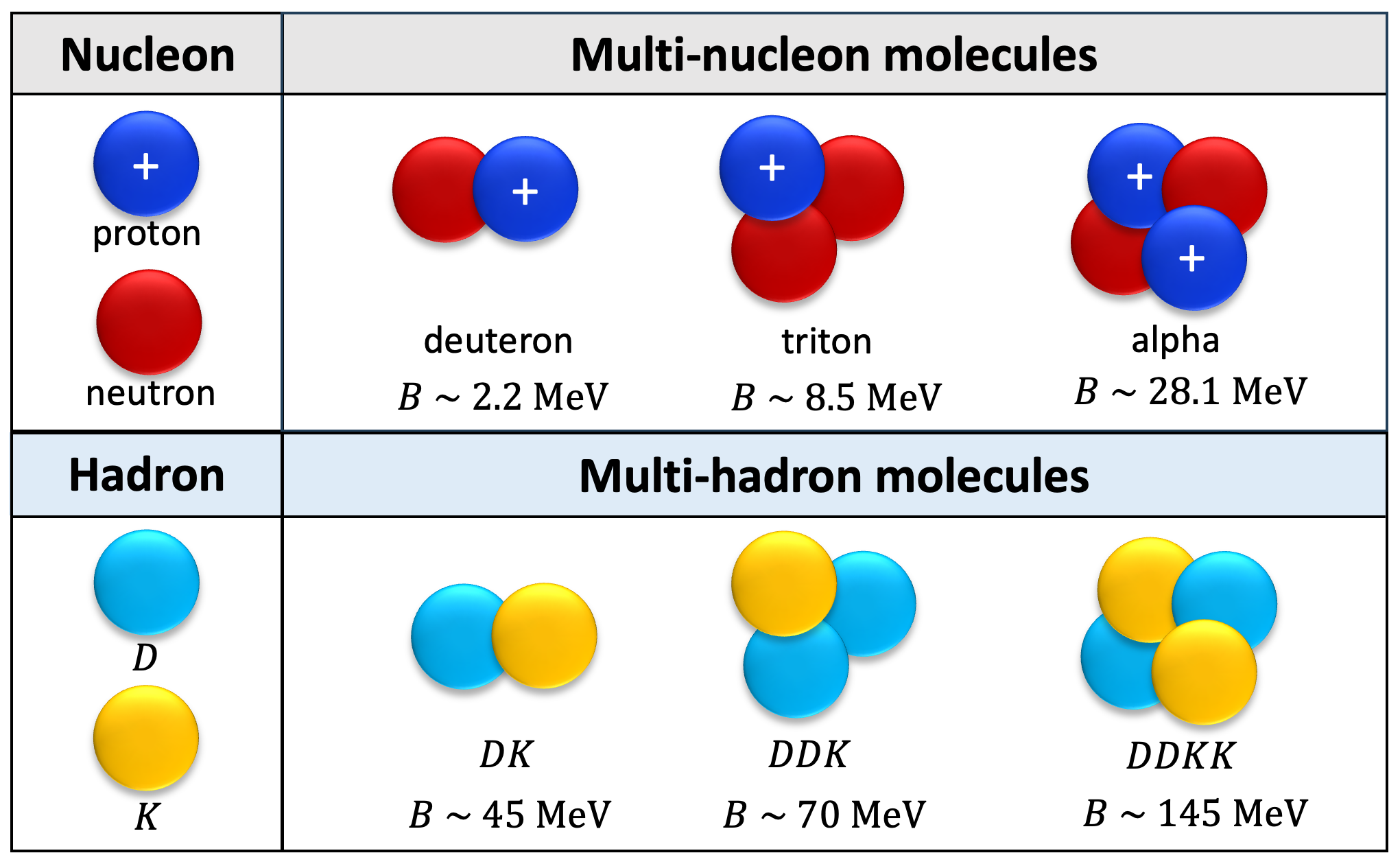}
\end{overpic}
\caption{ Analogies between the few-body bound states of nucleons, deuteron, triton, and alpha, and the few-body bound states of $D$ and $K$. 
}
\label{multi}
\end{center}
\end{figure}

In this work, we studied the four-body systems with $I(J^{P})=0(0^{+})$ composed of two $D$ mesons and two kaons (antikaons), $DDKK$ and $D\Bar{D}K\Bar{K}$. 
We adopted the OBE model to describe the $DD$ potential, and the cutoff is determined by reproducing the $X(3872)$ as $D\bar{D}^*$ molecule. We employed the WT term as the LO chiral potential and a repulsive core as the NLO correction in the non-relativistic limit to describe the $S$-wave $DK$ interaction. The uncertainties of the $DK$ interaction were estimated by varying the cutoff $R_{C}$ and coupling constant $C'_{S}$. The other coupling constant $C'_{L}$ was determined by reproducing the  $D_{s0}^{*}(2317)$ as a $DK$ bound state.  The strength of the $D\Bar{K}$ interaction was taken as half of the $DK$ interaction according to chiral perturbation theory. We employed a non-relativistic form to describe the interaction between two kaons and that between a kaon and an anti-kaon, which were expressed in the form of one Gaussian function. Two interaction ranges were considered. The strengths of the $KK$ and $K\Bar{K}$ interaction were obtained by reproducing the scattering length $a_{K^{+}K^{+}}=-0.14$ and fitting to the masses and widths of $f_{0}(980)$ and $a_{0}(980)$, respectively.

The four-body Schr\"{o}dinger equations were solved using the Gaussian expansion method. For the $DDKK$ system, the binding energy is about $138 \sim 155$ MeV. The dominant contribution is the $DK$ interaction, and the strength of the repulsive $KK$ interaction is compatible with the strength of the attractive $DD$ interaction, which has a tiny impact on the binding energy. As a result, we obtained a complete multi-hadron picture composed of $D$ mesons and kaons similar to that of nucleons, as shown in Fig.~\ref{multi}. The $DK$ molecule corresponds to deuteron ($np$), $DDK$ to triton ($nnp$), and $DDKK$ to the alpha particle ($nnpp$).

Although, the only difference between the $DD\Bar{K}\Bar{K}$ system and the $DDKK$ system is the interaction between the $D$ meson and the antikaon (kaon), the $DD\Bar{K}\Bar{K}$ system can not bind because the $\bar{D}K$ system does not bind. 
The $D\Bar{D}K\Bar{K}$ system is a bit more complicated due to the nonexistence of identical Jacobian coordinates and the imaginary term of the Hamiltonian. We treated the imaginary term perturbatively based on the wave functions obtained with only the real part of the Hamiltonian. The binding energy is $123 \sim 163$ MeV, and the decay width is $37 \sim 53$ MeV. The perturbative treatment is justified via $|\langle\Psi|\rm{Im}V|\Psi\rangle|\ll |\langle\Psi|\rm{Re}V|\Psi\rangle|$.

Undoubtedly, experimental searches for and further theoretical studies of these four-hadron molecules are essential to verify the molecular nature of the $D_{s0}^*(2317)$ and test our understanding of the $DK$, $D\bar{K}$, $DD$, $KK$, and $K\bar{K}$ interactions. According to Ref.~\cite{Wu:2022wgn}, the prompt production rates of these multi-hadron molecules in $e^+e^-$ colliders might be too small to be realistic. One needs to study other processes, such as heavy-ion collisions, to search for these states.

\section{Acknowledgments}
This work is partly supported by the National
Natural Science Foundation of China under Grants 
No.11975041, No.11961141004, and the National Key R\&D Program of China under Grant No. 2023YFA1606700. Ming-Zhu Liu acknowledges support from the National Natural Science Foundation of China under Grant No.12105007.  Junxu Lu acknowledges support from the National Natural Science Foundation of China under Grant No.12105006.

\bibliography{DDKK.bib}

\end{document}